\documentstyle[aps,prl,multicol,epsf]{revtex}

\begin{document}
%\draft
\title{Orbitally Driven Spin Pairing in the 3D   
Non-Magnetic Mott Insulator BaVS$_3$: Evidence from Single Crystal Studies}
\author{
 G. Mih\'{a}ly$^{1,2}$, I. K\'{e}zsm\'{a}rki$^{2}$, 
 F. Z\'{a}mborszky$^{2}$, M. Miljak$^{3}$, K. Penc$^{4,5}$, 
 P. Fazekas$^{2,5}$, H. Berger$^{1}$ and L. Forr\'{o}$^{1}$}

\address{
$^{1}$Institut de G\'eniue Atomique, Ecole Politechnique 
Federale de Lausanne, CH-1015 Lausanne, Switzerland\\
$^2$Department of Physics, Technical University of Budapest, H-1111 Budapest, 
Hungary \\
$^3$Institute of Physics of the University, P.O.Box 304, Zagreb, Croatia\\
$^4$Service de Physique Th\'eorique, C.E.A. Saclay, 91191 Gif-sur Yvette
Cedex, France\\
$^5$Research Institute for Solid State Physics and Optics, H-1525 Budapest, 
P.O.B. 49, Hungary }
\date{\today}
\maketitle

\begin{abstract}
Static electrical and magnetic properties of single crystal BaVS$_{3}$ were 
measured over the structural ($T_{S}=240$K), metal--insulator 
($T_{\rm MI}=69$K), and suspected orbital ordering ($T_{X}=30$K) 
transitions. The resistivity is almost isotropic both in the metallic 
and insulating states. An anomaly in the magnetic anisotropy at $T_{X}$ 
signals a phase transition to an ordered low-T state. The results are 
interpreted in terms of orbital ordering and spin pairing within the 
lowest crystal field quasi-doublet. The disordered insulator 
at $T_{X}<T<T_{\rm MI}$ is described as a classical liquid of 
non-magnetic pairs.
\end{abstract}

\begin{multicols}{2}
\narrowtext

Spatial ordering of the occupancy of degenerate electronic orbitals plays
important role in the diverse magnetic phenomena of transition metal compounds 
\cite{LN}. To cite a well-known example: the interplay of magnetic and 
orbital long range ordering, and strong coupling to the lattice, account  
for the metal--insulator transitions of the V$_{2}$O$_{3}$ system 
\cite{Bao,Paolasini}. In contrast, the metal--insulator transition of the 
$S=1/2$, $3d^{1}$ electron system BaVS$_{3}$ is not associated either with 
magnetic long range order, or with any detectable static spin pairing. As an 
alternative, the possibility of an orbitally ordered ground state was 
discussed \cite{Nakamura}, while other proposals emphasized the 
quasi-one-dimensional character of the material \cite{Graf,Nakamura2,Imai}.
The crystal structure is certainly suggestive of a linear chain compound 
since along the $c$ axis, the intrachain V--V distance is only 2.81 \AA, 
while in the $a$--$b$ plane the interchain separation is 6.73 \AA 
\cite{Gardner,Ghedira}. It is thus somewhat surprising that our present 
studies show that electrically BaVS$_{3}$ is nearly isotropic. This means 
that BaVS$_{3}$ provides one of the few realizations of a Mott 
transition within the non-magnetic phase of a three-dimensional system. Since 
this case (or rather its $D\to\infty$ counterpart) is much studied 
theoretically, but scarcely investigated experimentally, a good understanding 
of BaVS$_3$ should be valuable for strong correlation physics in general. 

BaVS$_{3}$ has a metal-insulator transition at $T_{{\rm MI}}=69$K,
accompanied by a sharp spike in the magnetic susceptibility \cite
{Graf,Massenet}. The high temperature phase is a strongly correlated metal
with mean free path in the order of the lattice constant. There is no sign
of a sharp Fermi-edge in the UPS/XPS spectra \cite{Nakamura2} and instead of
a Pauli-susceptibility it exhibits Curie-Weiss like behavior. Though the
magnetic susceptibility is similar to that of an antiferromagnet \cite
{Massenet,Takano}, no long-range magnetic order develops at the transition 
\cite{Ghedira,Heidemann}. The transition is clearly seen in the 
thermal expansion anomaly \cite{Graf}, and in the specific heat \cite{Imai}. 
The $d$-electron entropy right above $T_{\rm MI}$ is estimated as 
$\sim 0.8 R \ln{2}$, and it seems that a considerable fraction of the 
electronic degrees of freedom is frozen even at room temperature \cite{Imai}. 
It appears that the 69K transition is not symmetry breaking\cite{bignote}: 
it is a pure Mott transition which does not involve either 
magnetic order or any static displacement of the atoms.

Hints of long range order were found well below $T_{{\rm MI}}$, at 
$T_{X}=30$K, in recent NMR and NQR experiments \cite{Nakamura}. It was 
suggested that orbital order may develop here, but it could not be decided 
whether the $T_X$ phenomenon is a true phase transition, or purely dynamical.
In any case, the associated entropy change must be very small since it escaped 
detection \cite{Imai}. In this work we prove that there is a phase 
transition, by finding its signature in static magnetic properties.

In order to clarify the nature of BaVS$_{3}$,  we performed single
crystal experiments, searching for macroscopic anisotropy in the electrical 
and the magnetic properties. Our results exclude the quasi-1D 
interpretations and supply direct evidence for the dominant role of 
$e(t_{2g})$ orbitals tilted out from the chain direction. The static 
magnetic susceptibility $\chi$, and more markedly, its anisotropy 
$\chi_c-\chi_a$, show clear anomalies both at $T_{\rm MI}$ and $T_X$. 

Single crystals of BaVS$_{3}$ were grown by Tellurium flux method.
Powders of BaVS$_{3}$ and sublimated tellurium 99.99\% Ventron were mixed in
a molar ratio 1 : 50 and heated up to 1050 $^{o}$C in an evacuated silica
ampoule. Then it was slowly cooled down to $55 ^{o}$C at the rate of 
$1^{o}$C/hour. The crystals, obtained from the flux by sublimation, have 
typical dimensions of $3 \times 0.5 \times 0.5 $ mm$^{3}$.

For the longitudinal ($c$-direction) resistance measurement standard four
probe contact arrangement was applied. The conduction anisotropy was
determined by the Montgomery method \cite{Montgomery}. Experiments performed on
several crystals from various preparation batches showed deviation only at
low temperatures, which we attribute to the different purity of the samples
(see later). The magnetic susceptibility was measured by Faraday-balance,
the anisotropy measurements were carried out on a sensitive torque
magnetometer \cite{Marko}.

\vspace{-0.3truecm}
\begin{figure}[tbp]
 \epsfxsize=7.0 truecm
 \centerline{\epsffile{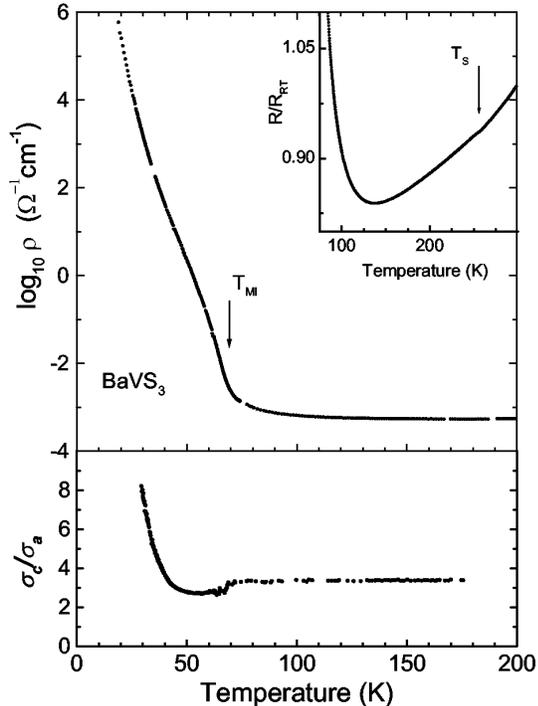}} 
\caption{Temperature dependence of the resistivity ${\protect\rho }_{c}(T)$, 
and the conduction anisotropy $\protect\sigma _{c}/\protect\sigma _{a}$ in
BaVS$_{3}$. The arrows indicate $T_{S}$, $T_{{\rm MI}}$ and $T_{X}$,
respectively. }
\label{fig1}
\end{figure}
  
  Figures 1 and 2 show the temperature dependence of the resistivity and the
conduction anisotropy of single crystals. BaVS$_{3}$ is a ``bad
metal'' with $\rho _{{\rm RT}}=0.7$m${\Omega}$cm. With decreasing
$T$ a slight change of slope reflects the structural transition at 
$T_{S}=240$K (Fig. 1, upper panel). The resistivity has a minimum at 
$125$K. A sharp metal-insulator transition sets in at $T_{\rm MI}=69$K 
(Fig.~2 inset).
Below $T_{\rm MI}$ the resistivity increases steeply and it varies 9 
orders of magnitude down to about 20K. Crossing $T_{X}$ does not influence 
$\rho(T)$ as it is obvious also from the Arrhenius plot (Fig. 2). We deduce 
a gap $\Delta \approx 600$K for the insulator ($\Delta$ is not well defined 
due to a slight curvature in the $\ln{\rho} - 1/T$ plot). The overall 
behavior of $\rho (T)$ agrees well with that of high purity polycrystalline 
samples \cite{Graf}.

The conduction anisotropy, defined as the ratio of conductivities measured 
along and perpendicular to the chain direction, is surprisingly low, 
$\sigma_{c}/\sigma _{a}\approx 3$. It is temperature independent in the 
metallic phase and there is only a small jump at the metal--insulator 
transition. Below $T_{{\rm MI}}$ the anisotropy has the same small value
over a broad temperature range down to about $30-40$K. Note that in this
range of $T$ the resistivity increases about 6 orders of magnitude in
both directions. The low temperature upturn is related to impurities, the
two most different results obtained on different samples are plotted in 
Fig. 2 (lower panel). It seems that impurity donated carriers enhance the 
$c$-direction conduction, and for this process the activation energy is about 
$70$K.

\vspace{-0.3truecm}
\begin{figure}[tbp]
 \epsfxsize=7.0 truecm
 \centerline{\epsffile{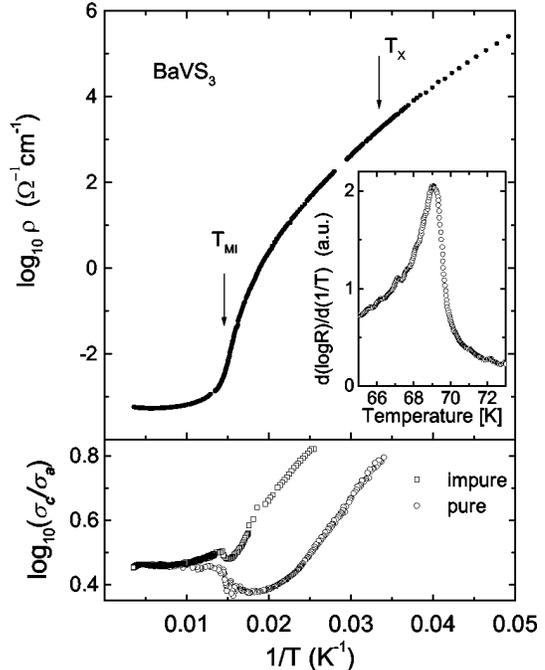}}
\caption{Arrhenius plot of the resistivity and the conduction anisotropy.
The latter is shown for two crystals of different purity. 
Inset: the peak of $d\ln \rho/d (1/T)$ defines $T_{\rm MI}$.}
\label{fig2}
\end{figure}

Figure 3 summarizes the results of the magnetic susceptibility measurements.
Along the chain direction, $\chi_{c}(T)$  agrees well with previous data on
stoichiometric samples; the Curie--Weiss behavior in the metallic phase  
($\mu_{\rm eff}=1.2\mu_{B}$; the small $\Theta< 10$K may vary 
slightly with the range of fit) is followed by a steep decrease in the 
susceptibility below 
$T_{\rm MI}$.  $\chi_{c}$ and $\chi_{a}$ look very similar (therefore we
show only $\chi_{c}$), and both are fairly smooth at $T_{X}$. However, we
find a sharp peak in the temperature derivative of the $a$-axis
susceptibility $d\chi _{a}/dT$, and a sudden break in the anisotropy 
$\chi_{c}-\chi_{a}$ at $T_{X}\approx 30$K. These give convincing evidence 
of a phase transition within the insulating phase. We suggest that the 
transition temperature $T_{X}$ be defined by the sharp peak in 
$d\chi_{a}/dT$ \cite{footnote3}.

Discussing the results first we interpret the low value of the conduction
anisotropy. For the present purposes, we assume that the simple ionic picture 
holds (Ba$^{2+}$V$^{4+}$S$_{3}^{2-}$ ionic state, $3d^{1}$ configuration). 
Figure 4a shows the
crystal field splitting of the vanadium $d$-levels. The sulphur octahedra 
surrounding nearest neighbor vanadium ions are face-sharing along the 
$c$-axis. Above $T_S=240$K, there is a trigonal distortion along the $c$-axis; 
for $T<T_S$, an additional orthorhombic component appears. The main effect 
is the trigonal splitting of the $t_{2g}$ level, which lifts the $d_{z^{2}}$ 
level (the $z$-axis being now the trigonal axis) above the two degenerate 
$e(t_{2g})$ orbitals \cite{Massenet}. The remaining degeneracy is lifted 
by the orthorhombic distortion: below $T_{S}$, the low-energy crystal field 
states can be thought of as a quasi-doublet with a small splitting.   

\vspace{-0.3truecm}
\begin{figure}[tbp]
 \epsfxsize=7.0 truecm
 \centerline{\epsffile{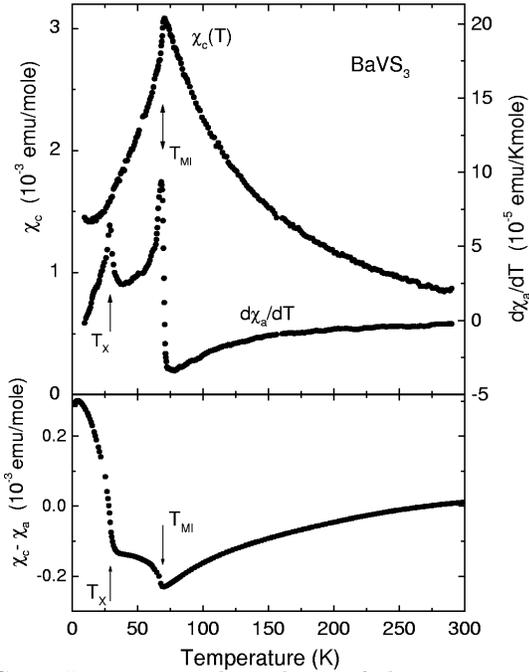}}
\caption{Temperature dependence of the $c$-axis magnetic susceptibility 
$\protect\chi_{c}$, the derivative of the $a$-axis susceptibility 
$d\protect\chi_a /dT$, and the susceptibility anisotropy, 
$\protect\chi _{c}-\protect\chi _{a}$.}
\label{fig3}
\end{figure}

In Fig. 4 the lobes of the low-lying orbitals are shown on the background of 
the crystal structure. The $d_{z^{2}}$ orbitals have large direct overlap along
the chain direction as shown in the left side of Fig. 4b. Taking also into
account that the vanadium chains are widely separated, any
conduction through the $d_{z^{2}}$ channel is expected to be extremely
anisotropic. In contrast, there is only weak, indirect overlap of the two 
$e(t_{2g})$ levels \cite{footnote2}, and hopping along these channels is
almost isotropic. From the observed small conduction anisotropy we conclude
that the $d_{z^{2}}$ orbitals are not involved in the electron transport. In
contrast to previous assumptions \cite{Massenet}, the crystal field spitting
between $d_{z^{2}}$ and the $e(t_{2g})$ levels must be large and the band
formed from the $d_{z^{2}}$ orbitals does not overlap with the occupied
states.

We believe that the electron propagation along the $c$-axis occurs also
through multiple nearest neighbor interchain hops involving only the 
$e(t_{2g})$ orbitals. Extended states of the $d_{z^{2}}$ band are excited
only at low temperatures from impurity levels situated below the broad band.
This process is seen in the impurity dependent low-temperature upturn of the
anisotropy.

The susceptibility data show that the number of polarizable moments
drops drastically upon entering the insulating phase. On the other hand, the
metal--insulator transition has no magnetic precursor on the metallic side,
$\chi$ closely follows the paramagnetic formula down to $T_{\rm MI}$. 
The volume change \cite{Graf} and the onset of an $ab$-plane
rearrangement of the V sublattice at the transition \cite{Ghedira} 
indicate that the state
of the system changes profoundly, and the effective spin--spin interaction 
may be completely different from that in the metallic phase. It seems that 
the insulator phase does not arise from a magnetic instability of the 
metallic phase. Let us recall the case of V$_{2}$O$_{3}$
where the onset of orbital order \cite{Paolasini} leads to an insulating 
state whose magnetic ordering pattern cannot be
anticipated from the short range order found in either of the neighboring
phases \cite{Bao}. We propose a similar picture for BaVS$_{3}$: the
metal--insulator transition involves both the spin and orbital degrees of
freedom, and considering only the effect on spins, it amounts to a change of
the spin hamiltonian. It is the consequence of the frustrated structure 
(a triangular array of V chains), and of the form of the relevant 
crystal field states, 
that the intermediate ($T_X<T<T_{\rm MI}$) phase is not an ordered magnet, 
but an overall non-magnetic state with peculiar spin and orbital short 
range order.

\vspace{-0.3truecm}
\begin{figure}[tbp]
 \epsfxsize=7.0 truecm
 \centerline{\epsffile{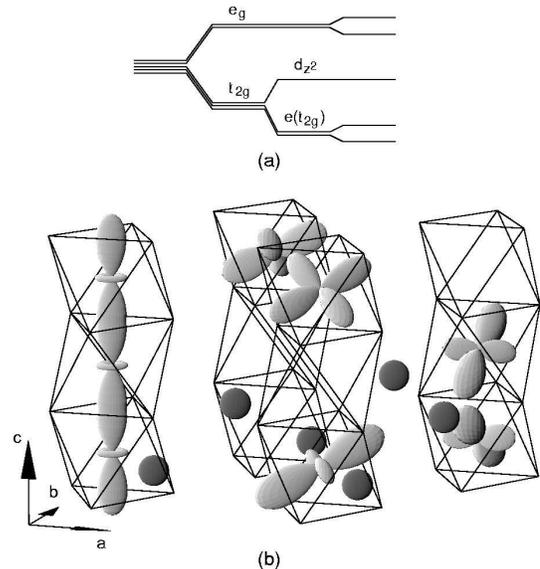}}
\caption{(a) Crystal field splitting of the vanadium $3d$ levels due to the
symmetry reduction from octahedral $\rightarrow $ trigonal (distortion along 
the $c$-direction) $\rightarrow $ orthorhombic (zig-zag chain developed below 
$T_{S}$). (b) Perspective view of the crystal structure of BaVS$_{3}$. The 
orientation of the low lying orbitals derived from $t_{2g}$ is also 
displayed.}
\label{fig4}
\end{figure}

In order to investigate the electronic structure of the insulating
phase, we derived a Kugel-Khomskii type model \cite{KKM} starting from the
atomic limit, including also the orbital index dependence of the hopping
matrix elements, and the spin--orbit coupling \cite{penc}. There is a 
broad range of parameters for which the ground state of an isolated pair of 
sites is non-magnetic. Further numerical and variational calculations for 
clusters of up to 24 sites revealed the presence of a large 
number of ordered low-energy states. Figure 5 gives examples of these 
various long-period structures which obey characteristic short range rules 
for the relative orientation of the pairs. 
Due to the small energy difference between 
the various configurations the singlet pair formation at $T_{{\rm MI}}$ is not 
accompanied by true long range order, but instead a spin--orbital 
liquid develops. Thermal averaging over the (exponentially large number of) 
accessible configurations gives
rise to a homogeneous state with low susceptibility. Moreover, in accordance
with NMR/NQR results \cite{Nakamura}, no static pairing is expected over a
broad temperature interval below $T_{{\rm MI}}$. Long range static order is
reached only at the much lower temperature $T_{X}$, as shown by the present 
susceptibility data, and also by anomalies in microscopic
measurements \cite{Nakamura,Sayetat}. Due to the pre-existing short range
order, this transition is not accompanied by any significant entropy change.

\vspace{-0.3truecm}
\begin{figure}[tbp]
 \epsfxsize=6.5 truecm
 \centerline{\epsffile{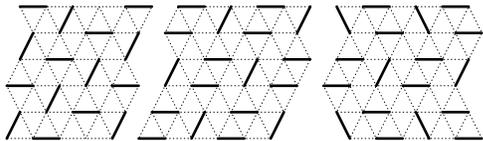}}
\caption{Some of the energetically favorable pair arrangements 
in the triangular $ab$ plane.}
\label{fig5}
\end{figure}

The pictures in Fig. 5 may remind us of the resonating valence bond state of 
frustrated Heisenberg models \cite{LN}. We emphasize that the situation is 
completely different here. 
Within each pair, those orbitals are occupied which give rise to a strong 
intrapair exchange, and thus each cluster state belongs to a different 
effective spin Hamiltonian. Considering the shape of the $e(t_{2g})$ orbitals 
one finds that pair formation does not quite saturate the exchange 
interaction, but there are residual interactions which govern 
farther-neighbor correlations, and cause a weak resonance. 
However, this resonance is much weaker than in 
the pure $S=1/2$ Heisenberg models, and the  $T_X<T<T_{\rm MI}$ phase of 
BaVS$_3$ is better described as a thermal average over valence bond solids, 
than as a resonating valence bond liquid. The finding of a 
$T$-dependent spin gap, which vanishes at $T_{\rm MI}$ \cite{Nakamura,spingap},
is consistent with our scenario. 

The metal--insulator transition affects mainly the spin degrees of
freedom: the electronic entropy present at $T$ slightly above $T_{\rm MI}$ 
\cite{Imai} is almost exhausted by the spin entropy required by 
the measured $T>T_{{\rm MI}}$ susceptibility which can be ascribed to 
$\sim 70$\% of the V sites carrying nearly independent localized spins. The 
primary order parameter of the low temperature phase is the density of the 
non-magnetic pairs, and the driving force of the transition is the gaining of 
spin entropy. Sizeable orbital order must exist even above the metal--insulator
transition (below $T_{{\rm MI}}$, it becomes additionally stabilized by 
the singlet formation as shown by the appearance of an extra component of
the orthorhombic distortion \cite{Sayetat}). 
We emphasize, however, that the entropy which would 
correspond to the complete absence of short range orbital order is certainly 
not present even at $T=300$K \cite{Imai}. 

In conclusion, we have shown that electron propagation in BaVS$_{3}$
occurs via nearest neighbor interchain hops involving the 
$e(t_{2g})$ orbitals only. The metal--insulator transition was described as a 
transition to a classical liquid of non-magnetic pairs, which shows 
spin {\sl and} orbital short range order. We have
proposed that though farther-neighbor inter-pair interactions prefer
ordered structures, due to the frustrated triangular structure long
range order can develop only well below the metal--insulator transition, at 
$T_{X}$. Our explanation is in overall agreement with the results of 
transport, magnetic and thermodynamic experiments.

This work was supported by the Swiss National Foundation for Scientific
Research and by the Hungarian Research Funds OTKA T025505, FKFP 0355, B10, and
AKP 98-66.

\end{multicols}

\begin{references}
\bibitem{bignote} The only known change of symmetry 
occurs at the hexagonal-orthorhombic transition at $T_{S}=240$K. 
It slightly affects the resistivity, and it is not visible 
either in the magnetic susceptibility or in the specific heat.
\bibitem{footnote3} The phase transition would be less visible in 
$d\chi_c/dT$ which has only a small bump.
\bibitem{footnote2}  The $e(t_{2g})$ orbitals shown in Fig. 4 are the $\pm 1$
eigenstates of the orbital angular momentum $L^{z}$: 
$ d_{x}=(x^{2}-y^{2})/\sqrt{6}-xz/ \sqrt{3}$ and 
$d_{y}=xy/\sqrt{3/2}+yz/\sqrt{3}$. 
Note that the coordinate system has beeen rotated from the 
octahedral main axes so that now $z$ is along the trigonal axis.
\end{references}
\end{document}